\documentclass[aps,prl,twocolumn,amsmath,amssymb,nofootinbib,superscriptaddress,floatfix,reprint,longbibliography]{revtex4-1}
\usepackage[latin1]{inputenc}
\usepackage{amsmath}
\usepackage{amsfonts}
\usepackage{amsthm}
\usepackage{amssymb}
\usepackage{lipsum}
\usepackage{graphicx}
\usepackage[usenames,dvipsnames]{color}
\usepackage{float}
\usepackage{multirow}
\usepackage{soul}
\usepackage{physics}
\usepackage{appendix}
\usepackage[colorlinks=true, urlcolor=blue, linkcolor=blue, citecolor=blue, pdftex]{hyperref}
\usepackage{dcolumn}
\usepackage{bm}
\usepackage{url}
\usepackage{braket}
\usepackage[range-units=single,separate-uncertainty]{siunitx}
\usepackage{verbatim}

\begin{document} 
	
\author{Armando Consiglio}\email{armando.consiglio@physik.uni-wuerzburg.de}\affiliation{Institut f\"{u}r Theoretische Physik und Astrophysik and W\"{u}rzburg-Dresden Cluster of Excellence ct.qmat, Universit\"{a}t W\"{u}rzburg, 97074 W\"{u}rzburg, Germany}

\author{Tilman Schwemmer}\affiliation{Institut f\"{u}r Theoretische Physik und Astrophysik and W\"{u}rzburg-Dresden Cluster of Excellence ct.qmat, Universit\"{a}t W\"{u}rzburg, 97074 W\"{u}rzburg, Germany}

\author{Xianxin Wu}\affiliation{Max-Planck-Institut f\"ur Festk\"orperforschung, Heisenbergstrasse 1, D-70569 Stuttgart, Germany}

\author{Werner Hanke}\affiliation{Institut f\"{u}r Theoretische Physik und Astrophysik and W\"{u}rzburg-Dresden Cluster of Excellence ct.qmat, Universit\"{a}t W\"{u}rzburg, 97074 W\"{u}rzburg, Germany}

\author{Titus Neupert} \affiliation{Department of Physics, University of Zurich, Winterthurerstrasse 190, 8057 Zurich, Switzerland}

\author{Ronny Thomale}\affiliation{Institut f\"{u}r Theoretische Physik und Astrophysik and W\"{u}rzburg-Dresden Cluster of Excellence ct.qmat, Universit\"{a}t W\"{u}rzburg, 97074 W\"{u}rzburg, Germany}\affiliation{Department of Physics and Quantum Centers in Diamond and Emerging Materials (QuCenDiEM) group, Indian Institute of Technology Madras, Chennai 600036, India}

\author{Giorgio Sangiovanni}\affiliation{Institut f\"{u}r Theoretische Physik und Astrophysik and W\"{u}rzburg-Dresden Cluster of Excellence ct.qmat, Universit\"{a}t W\"{u}rzburg, 97074 W\"{u}rzburg, Germany}

\author{Domenico Di Sante}\affiliation{Department of Physics and Astronomy, Alma Mater Studiorum, University of Bologna, 40127 Bologna, Italy}\affiliation{Center for Computational Quantum Physics, Flatiron Institute, 162 5th Avenue, New York, New York 10010, USA}

\title{Van Hove tuning of AV$_3$Sb$_5$ kagome metals under pressure and strain}

\date{\today}

\begin{abstract}
From first-principles calculations, we investigate the structural and electronic properties of the kagome metals AV$_3$Sb$_5$ (A = Cs, K, Rb) under isotropic and anisotropic pressure. Charge ordering patterns are found to be unanimously suppressed, while there is a significant rearrangement of $p$-type and $m$-type van Hove point energies with respect to the Fermi level. Already for moderate tensile strain along the V plane and compressive strain normal to the V layer, we find that a van Hove point can be shifted to the Fermi energy. Such a mechanism provides an invaluable tuning knob to alter the correlation profile in the kagome metal, and suggests itself for further experimental investigation. It might allow to reconcile possible multi-dome superconductivity in kagome metals not only from phonons, but also from the viewpoint of unconventional pairing.
\end{abstract}

\maketitle 
 
The recently discovered family of nonmagnetic quasi-2D AV$_3$Sb$_5$ (A = Cs, K, Rb) Kagome metals \cite{PhysRevMaterials.3.094407,PhysRevLett.125.247002} represents an excellent example of compounds that allow to study charge ordering \cite{Jiang2021UnconventionalCC,PhysRevB.87.115135, PhysRevLett.110.126405,song2021competition, ratcliff2021coherent, denner2021analysis, uykur2021optical, nakayama2021multiple, lou2021chargedensitywaveinduced,PhysRevB.104.035142,mielke2021timereversal}, superconductivity \cite{PhysRevB.86.121105, PhysRevB.79.214502, PhysRevMaterials.5.034801, wang2021enhancement, qian2021revealing,lin2021kagome} and electronic correlations \cite{PhysRevB.103.L241117}, together with Dirac band crossing, $\mathbb{Z}_2$ non-trivial topological bands \cite{setty2021electron}, chiral
symmetry breaking \cite{ciola2021chiral} and flat-band physics \cite{PhysRevMaterials.5.034801, ye2021flat}. 

Crystallized in the $P6/mmm$ space group, the compounds are based on a layer of Vanadium atoms arranged in a Kagome lattice and coordinated by Antimony atoms which are organized in two sublattices. One Sb sublattice exhibits a graphite-like structure sandwiching the V Kagome layer, while the other sublattice is formed by a single Sb atom centered inside the Kagome hexagon. 


In a Kagome lattice there are 3 sublattice sites per unit cell, giving rise to three electronic bands of which two are dispersive featuring a saddle point, {\it i.e.} van Hove singularity (vHs) at \textit{\textbf{M}}, while the remaining one is exactly flat assuming only nearest neighbor hybridization. Interestingly, at the \textit{\textbf{M}} points, the nesting momenta are commensurate and half the length of a reciprocal lattice vector. This leads to a 2$\times$2 enlarged unit cell in real space for translation symmetry breaking order. Recently, the different sublattice structures of the vHs for the Kagome lattice have been investigated \cite{wu2021nature}. The vHs of the upper dispersive kagome band is formed by eigenstates that feature only the contribution from one single sublattice site ($p$-type), contrary to the lower vHs where the eigenstates distribute over two sublattices ($m$-type) \cite{kang2021twofold,hu2021rich}. The $p$-type vs. $m$-type property of the vHs gives rise to different types of instabilities, and highlights the relevance of the substructure of a given vHs along with its proximity to the Fermi level $E_F$.

In this Letter, we intend to investigate the van Hove Fermiology profile of kagome metals as a function of pressure and strain. In AV$_3$Sb$_5$, the correlated phases that are experimentally observed depend not only on the thickness and the temperature of the material, but also on the applied external stress \cite{song2021competing, PhysRevB.103.L220504}. In particular, a departure from ambient pressure not only unfolds a superconducting dome descending from a charge density order parent state, but also yields a second dome feature of superconductivity for increasing pressure \cite{zhao2021nodal, PhysRevLett.126.247001, PhysRevB.103.224513, zhu2021doubledome}.
The possibility of tuning interactions and competition of phases using a thermodynamic quantity like pressure, rather than the more invasive introduction of chemical impurities, is indeed a powerful way to modify the Fermiology of a material. In the process of increasing pressure values, also the distance among atoms is modified, resulting in shorter chemical bonds and different electronic structures. We demonstrate how pressure, and in particular non-hydrostatic stresses \cite{tsirlin2021role}, lead to interesting effects in Kagome AV$_3$Sb$_5$, which may ultimately enhance the superconducting pairing strength.

Previous studies concerning structural instabilities in the AV$_3$Sb$_5$ compounds revealed the appearance of a 2$\times$2 super-lattice modulation \cite{Jiang2021UnconventionalCC, PhysRevX.11.031026}. It has also been pointed out that the inverse Star of David (ISD) deformation is favored by a 2$\times$2 Charge Density Wave (CDW) in all three cases (A = Cs, K, Rb) \cite{PhysRevLett.127.046401}, being locally stable and reducing the total energy with respect to the pristine phase. The same considerations also apply for the Star of David (SD) configuration, which is, however, less energetically favorable than the ISD one (see Supplementary Material~\cite{SuppMat}). Experiments have already started to explore the modification of the CDW order as a function of pressure \cite{wang2021competition,Yu_2021}, or the nature of the CDW itself \cite{wang2021origin}. First experiments also revealed the competition between CDW and superconductivity through uni-axial strain~\cite{PhysRevB.104.144506}.

\begin{figure}[tb]
	\centerline{\includegraphics[angle=0,width=0.85\columnwidth]{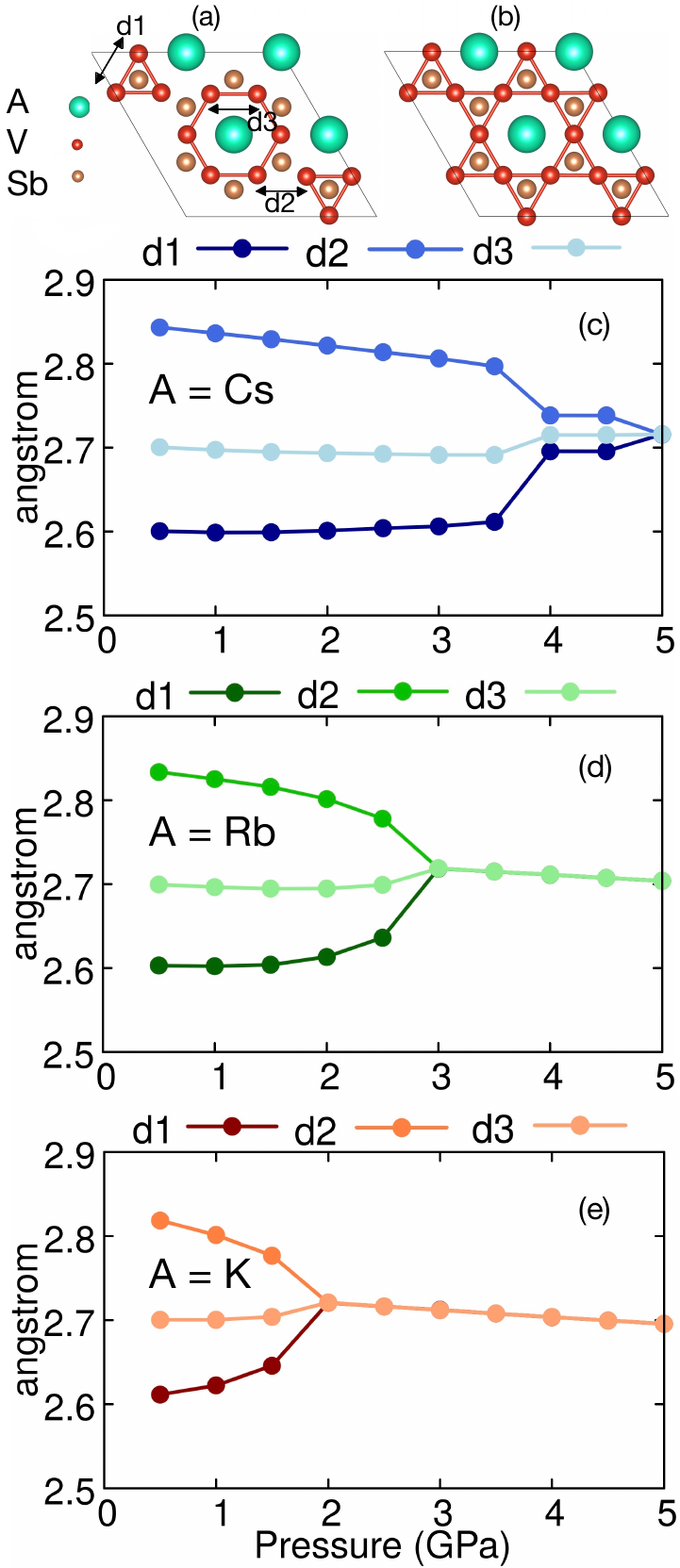}}
	\caption{Examples of ISD phase ($P = 0.5$ GPa, Panel \emph{a}) and pristine phase ($P \approx 4$ GPa, Panel \emph{b}) for the AV$_3$Sb$_5$ (A = Cs, K, Rb) Kagome metals. Panel \emph{c}: CsV$_3$Sb$_5$; Panel \emph{d}: RbV$_3$Sb$_5$; Panel \emph{e}: KV$_3$Sb$_5$. $\emph{d1}$ is the V-V distance inside the triangles, $\emph{d2}$ is the nearest V-V distance between the edge of a triangle and the edge of an hexagon, and $\emph{d3}$ is the nearest V-V distance inside the hexagons.\label{fig:ISDpristine}}
\end{figure}

\begin{figure*}[tb]
	\includegraphics[width=\textwidth, trim={2.6cm  20.8cm  2.6cm  2.6cm},clip]{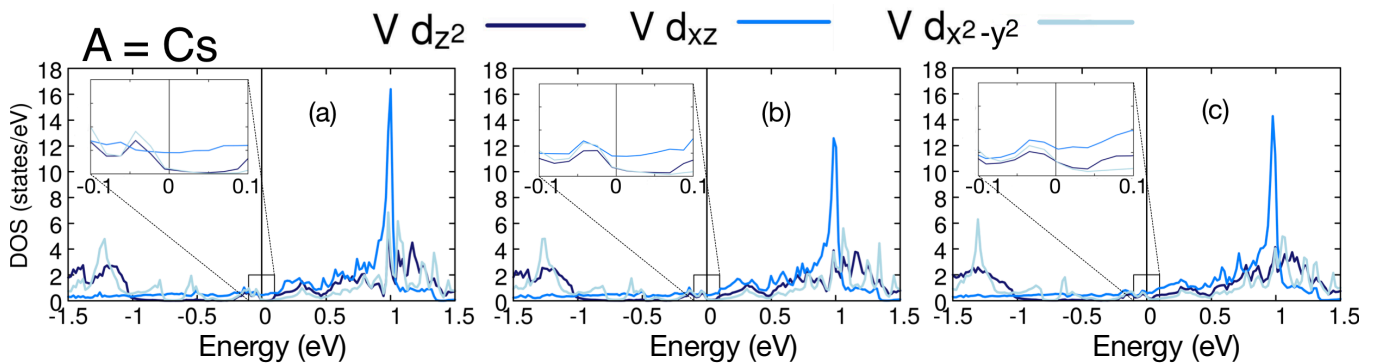}
	\caption{Normalized Orbital Resolved DOS for Vanadium $d$-orbitals in the ISD configuration of CsV$_3$Sb$_5$, at pressures $P =$ 0.5 (a), 1.5 (b) and 3.0 (c) GPa.}\label{fig:ISDDOS}
\end{figure*}

Starting from the ISD phase (Fig. \ref{fig:ISDpristine}\emph{a}), we gradually increased the applied external pressure monitoring the evolution of the orbital resolved Vanadium $d$-states close to $E_F$ (Fig. \ref{fig:ISDDOS}), as well as the bond lengths inside the Vanadium network (Figs. \ref{fig:ISDpristine} \emph{c, d, e}). Based on the assumption that pressures above a certain threshold will anyway result in a suppression of the charge ordering, we have not considered a detailed analysis about its true nature.
The CDW phase indeed tends to noticeably reduce or suppress vHs and the DOS in general at $E_F$, because CDW instabilities are related with softening of phonon modes and Fermi surface nesting \cite{PhysRevLett.2.393}. We found a rapid suppression of the ISD phase as a function of pressure, reaching the pristine 1$\times$1 phase after a certain threshold dependent on the chosen cation A. The lowest threshold value has been obtained for the KV$_3$Sb$_5$ Kagome metal (Fig. \ref{fig:ISDpristine} \emph{e}), possibly due to the shortest inter-layer distance among all considered compounds.


The evolution of the DOS as a function of pressure is shown in Fig.~\ref{fig:ISDDOS} for the case of CsV$_3$Sb$_5$. While the qualitative overall distribution is only weakly dependent on pressure, a closer inspection of the low-energy region around $E_F$ reveals an orbital dependent behavior (see Fig.~\ref{fig:ISDDOS}); the coexistence of CDW and superconductivity phases for small-enough pressures can be attributed to this orbital dependence. For $P = $ 0.5 GPa, the spectral weight with Vanadium $d_{z^2}$ and $d_{x^2-y^2}$ characters is vanishing while the $d_{xz}$ one remains finite. However, this scenario changes starting from $P = $ 3.0 GPa, with the Vanadium $d_{z^2}$ and $d_{x^2-y^2}$ characters becoming non-negligible at $E_F$. The cases A = K, Rb follow the same trend of the V-V bond lengths shown in Fig.~\ref{fig:ISDpristine}, with the ISD-pristine transition taking place for smaller pressure values (see also Supplementary Material~\cite{SuppMat}).

\begin{figure}[b]
	\includegraphics[width=\columnwidth]{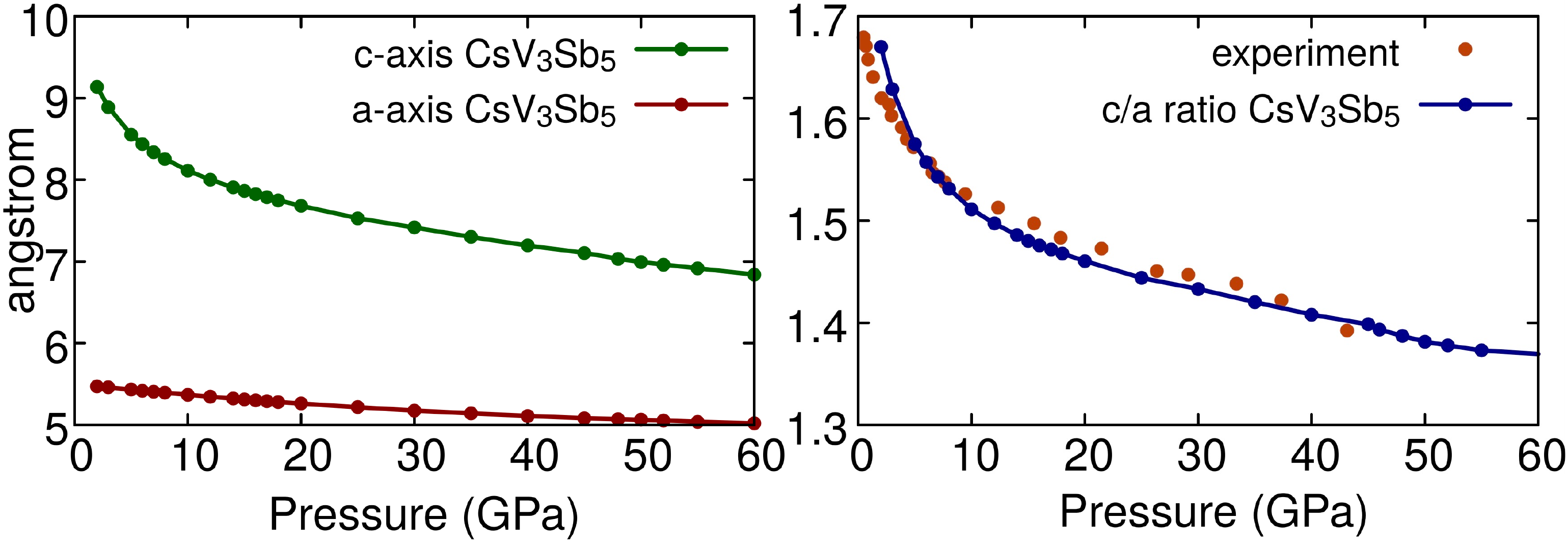}
	\caption{Left panel: $a$ and $c$ lattice constants as a function of pressure for CsV$_3$Sb$_5$. Right panel: $c/a$ ratio as a function of pressure; note also the slight change of slope for $P \simeq $ 45 GPa. The experimental data in right panel are taken from Ref.~\cite{PhysRevB.103.224513}. }\label{fig:latticeconstant_ratio}
\end{figure}

One of the most evident features upon increasing pressure is the strong compressive anisotropy, with the $c$-lattice parameter that tends to shrink faster than the $a$-lattice one, as a function of pressure (see Fig. \ref{fig:latticeconstant_ratio} for the specific case of CsV$_3$Sb$_5$). Also, while the $c$-lattice constant exhibits a highly non-linear behavior, the $a$-lattice constant is almost linear. The compression-isotropic scenario observed for relatively high pressure values is given by a reduced directional bonding. This is accompanied by a gradual formation of V$_3$Sb$_5$'s three-dimensional structures through bonding between the V$_3$Sb$_5$ slabs, intuitively leading to an increased dispersion along the $c$-axis. Pressure also leads to the formation of Sb$_2$-Sb$_2$ bonds. Finally, in agreement with experimental findings \cite{PhysRevB.103.224513}, we find that the $c/a$ ratio exhibits a change of slope for $P \approx 8$ GPa. This is worth noting because it allows to distinguish the low and high pressure regimes, but also because Lifshitz transitions can be associated to an anomaly in the $c/a$ ratio \cite{PhysRevB.97.115165}.

\begin{figure}[h]
\centerline{\includegraphics[angle=0,width=0.85\columnwidth]{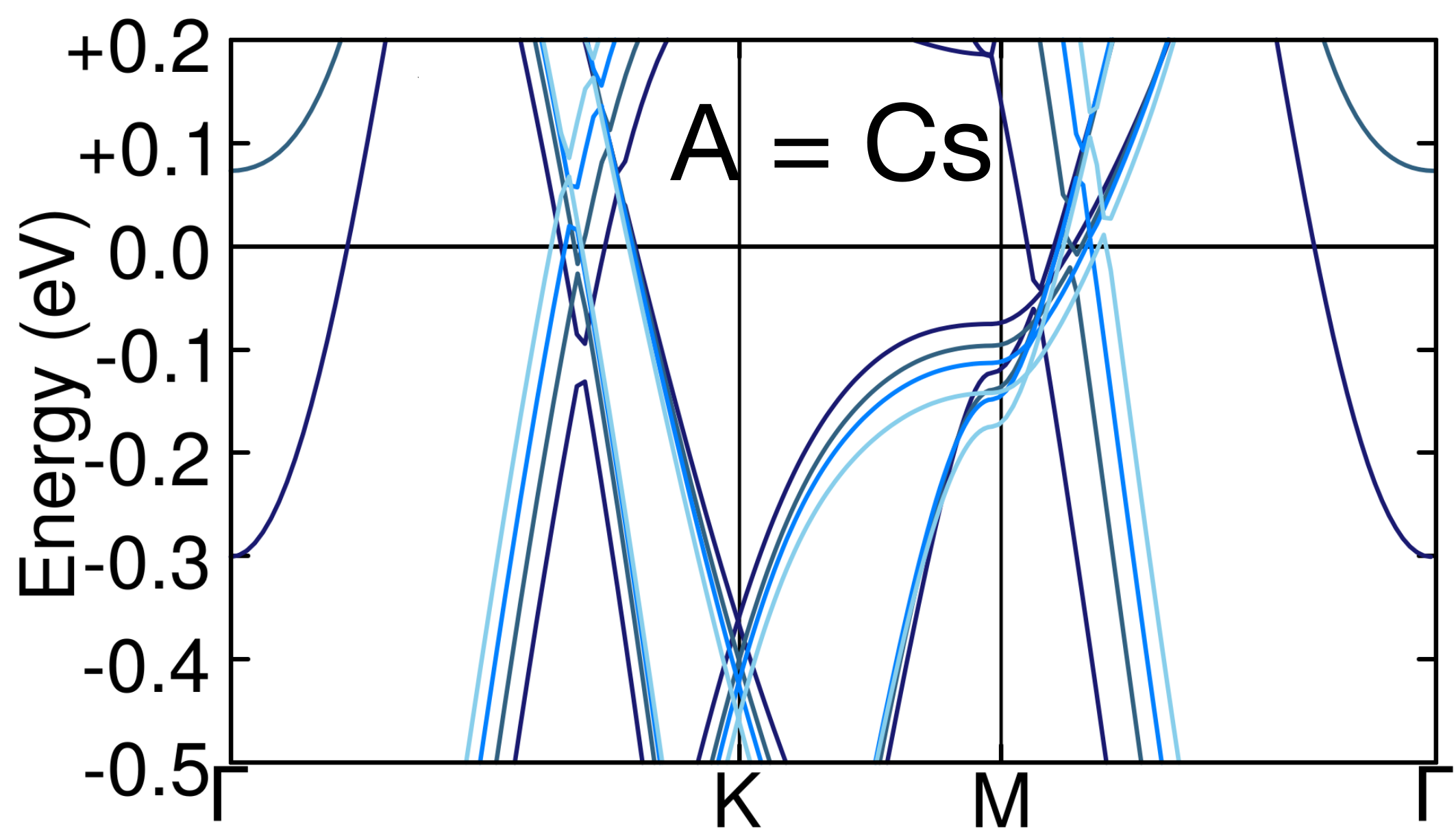}}
\caption{Evolution of electronic band structures as a function of pressure in the proximity of the Fermi level. The fading of colors, from darker to lighter tones, corresponds to a pressure increasing. Plotted bands are for $P = $ 3 GPa, 10 GPa, 15 GPa and 20 GPa.\label{fig:vhs}}
\end{figure}

Focusing now on the electronic band structure of AV$_3$Sb$_5$, the Antimony $p_z$ orbitals contribute primarily around the $\Gamma$ point, while the Fermi Surface close to \textit{\textbf{M}} is mainly due to the Vanadium $d$ orbitals. For all compounds, the general trend as a function of applied pressure, when this is high enough, is a decrease of the DOS due to the reduced inter-atomic distances, an obvious consequence of the increase of the bandwidth. From Fig.~\ref{fig:vhs} it can be noted how the saddle points at \textit{\textbf{M}} in proximity of $E_F$ tend to be pushed downward, away from $E_F$, upon increasing the pressure. A similar scenario (not shown) holds for $k_z = \pi/c$ as both vHss tend to be moved away from $E_F$, with the difference that now one vHs is above and the other below $E_F$. The $p_z$ projected bands experience the most visible changes with pressure, as seen in Fig.~\ref{fig:vhs}, and cross the Fermi level for $P \approx$ 7.5 GPa. On the contrary, the Sb $p_x$ and $p_y$ bands exhibit a much weaker modification, besides the aforementioned bandwidth broadening. Finally, the energy position of the two vHss for A = Cs is reversed with respect to A = K, Rb. Since the two vHss have different $d$-orbital characters, one can expect different Fermi surface properties, such as nesting features, for the CsV$_3$Sb$_5$ compound compared with the RbV$_3$Sb$_5$ and KV$_3$Sb$_5$ ones~\cite{kang2021twofold}.

Besides hydrostatic pressure, the electronic properties of materials can be tuned via an applied non-hydrostatic strain (compressive or tensile). An efficient method for applying a strain, for example, consists in growing the material on substrates with properly chosen in-plane lattice parameters. The difference (and advantage in some cases) of doing so compared with hydrostatic pressure is that, depending on the specific bulk modulus, it is possible to shrink some chemical bonds while increasing others. For the AV$_3$Sb$_5$ family, a tensile (compressive) strain along $a$ and $b$ tends to increase (decrease) the in-plane spacing among Vanadium atoms, and at the same time reduce (increase) the distance between the out-of-plane Antimony atoms in the unit cell.

\begin{figure}[tb]
	{\includegraphics[angle=0,width=0.85\columnwidth,trim={5.6cm  6.2cm  5.5cm  3cm},clip]{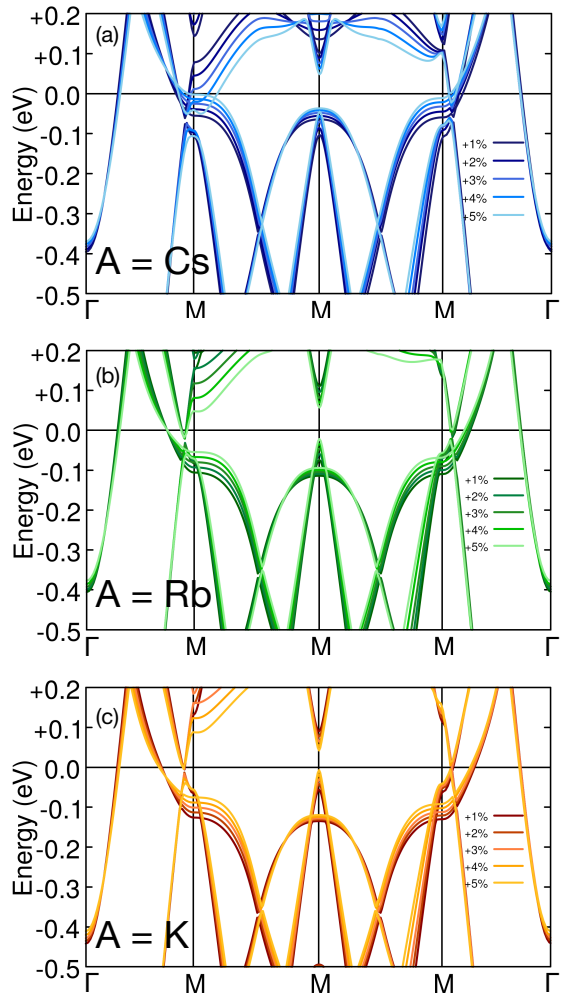}}
	\caption{Evolution of electronic band structures, with a focus on the 3 inequivalent \textit{\textbf{M}} points, as a function of uni-axial tensile strain along $a$-axis. The percentage numbers in the legend refer to the increase of $a$-axis. Note that these results have been obtained while keeping fixed the $c$-axis for all tensile and compressive strain values. Note that, for A = Cs, the vHs are swapped compared to the cases A = K, Rb. \label{fig:uniaxial_a}}
\end{figure}

Figure \ref{fig:uniaxial_a} shows the impact of uni-axial strain on the electronic properties of AV$_3$Sb$_5$. The main outcome of this analysis is represented by the vHs evolution: uni-axial tensile strain along $a$ is extremely effective in pushing the vHs closer to $E_F$. This is different from the effect of hydrostatic pressure described before. The relevance of our theoretical finding derives from recent experimental works which have reported the possibility of applying uni-axial pressures of up to $\sim$1 GPa, making use of a novel piezoelectric apparatus \cite{doi:10.1126/science.aaf9398}. This, in our case, would correspond to a uni-axial distortion along $a$ of $\sim$ 1\%. In the Supplementary Material~\cite{SuppMat} we show that in-plane bi-axial tensile strain, as well as uni-axial compressive strain along $c$, are likewise efficient means to tune a certain vHs closer to $E_F$.

In order to better disentangle the different contributions to the final result upon distorting the in-plane lattice parameters, we have kept the $c$-axis fixed. In a real experiment, the in-plane elongation of the lattice parameters leads to a concomitant negative relaxation of the out-of-plane lattice constant. In a first approximation justified under small external stress, this is naively understood from the conservation of the unit-cell volume. It means that, concerning the evolution of the electronic properties, one must take into account the combined effect of in-plane tensile strain and the connected compressive strain along $c$. As a result, the vHs will likely move closer to $E_F$ at experimentally accessible strain values.

{\it Conclusions:} Motivated by recent studies and results on the AV$_3$Sb$_5$ kagome metal compounds, we employed \emph{ab-initio} methods to monitor the evolution of the electronic and structural properties of this class of materials under a wide range of pressure values, both hydrostatic and non-isotropic. Starting from the ISD and SD charge ordered phases, in line with previous theoretical studies~\cite{labollita2021tuning, wang2021origin}, we observe for all three compounds (A = Cs, K, Rb) a suppression of the CDW already beyond a comparatively low pressure threshold ($P \approx 0.5 - 2.5$ GPa), also due to the anisotropic compression of the crystal structures. 
Upon higher hydrostatic pressure values ($P >$ 5 GPa), we observe a change around $E_F$ of the Sb $p_z$ electronic bands, leading to the disappearance of the pocket at the $\Gamma$ point. With regard to the Vanadium $d$ bands, which mainly contribute at the \textit{\textbf{M}} points, we find an increase of the overall bandwidth accompanied by a gradual shift of the saddle points away from $E_F$. In contrast, by applying uni- and bi-axial deformations, it is possible to efficiently move a vHs level in the opposite direction, {\it i.e.} closer to $E_F$. This is particularly evident upon considering compressive strain along the $c$-axis and/or tensile strain in the $a$-$b$ plane of Vanadium atoms. The distance between Antimony atoms and the Kagome net along $c$ hence assumes a primary role. 
As a consequence of a pressure-induced vHs shift closer to $E_F$, even from a viewpoint of unconventional pairing, the propensity for superconductivity could increase as a function of pressure. 
We therefore propose strain engineering as a preeminently suited tool to optimize the superconducting transition temperature in kagome metals. In Ref.~\cite{PhysRevB.104.144506} an increase of $T_c$ upon uni-axial compression along the $c-$axis has been experimentally reported, and attributed to the suppression of the CDW. In our analysis, we complement this line of reasoning by emphasizing the pressure-induced relative energy shift of the vHs (see Fig.~\ref{fig:uniaxial_a} and Supplementary Material~\cite{SuppMat}), which is strongly affected by both in-plane and out-of-plane uni-axial strain. Pressure-induced effects on $T_c$ in kagome metals can in principle result both from phonons, and concomitant structural transitions, as well as from unconventional pairing due to van Hove tuning. Which one turns out to be the dominant effect has to be individually analyzed experimentally.  

{\it Acknowledgments:} The DFT work was supported by the Deutsche Forschungsgemeinschaft (DFG, German Research Foundation) through Project-ID 258499086-SFB 1170 and by the W\"{u}rzburg-Dresden Cluster of Excellence on Complexity and Topology in Quantum Matter-ct.qmat Project-ID 390858490-EXC 2147.
The research leading to these results has received funding from the European Union's Horizon 2020 research and innovation programme under the Marie Sk{\l}odowska-Curie Grant Agreement No. 897276.
This project has received funding from the European Research Council (ERC) under the European Union's Horizon 2020 research and innovation programm (ERC-StG-Neupert-757867-PARATOP).
The authors acknowledge the Gauss Centre for Supercomputing e.V. for providing computing time on the GCS Supercomputer SuperMUC-NG at Leibniz Supercomputing Centre.

\bibliography{AV3Sb5.bib}

\end{document}